\documentclass[fleqn,usenatbib]{mnras}
\usepackage{newtxtext,newtxmath}
\usepackage[T1]{fontenc}

\DeclareRobustCommand{\VAN}[3]{#2}
\let\VANthebibliography\thebibliography
\def\thebibliography{\DeclareRobustCommand{\VAN}[3]{##3}\VANthebibliography}

\usepackage{graphicx}
\usepackage{epstopdf}
\usepackage{amsmath}
\usepackage{comment,xcolor}
\usepackage{hyperref}
\usepackage{gensymb}
\usepackage{xspace}
\usepackage{array}
\newcommand{\sbgs}{star-forming galaxies\xspace}
\newcommand{\sbg}{star-forming galaxy\xspace}

\title[EGMF constraints from UHECRs from local galaxies]{Extragalactic magnetic field constraints from ultra-high-energy cosmic rays from local galaxies}

\author[A. van Vliet et al.]{
Arjen van Vliet,$^{1}$\thanks{E-mail: arjen.van.vliet@desy.de}
Andrea Palladino,$^{1}$
Andrew Taylor$^{1}$
and Walter Winter$^{1}$
\\
$^{1}$Deutsches Elektronen-Synchrotron, Platanenallee 6, Zeuthen, Germany
}

\date{Accepted XXX. Received YYY; in original form ZZZ}

\pubyear{2021}

\begin{document}

\label{firstpage}
\pagerange{\pageref{firstpage}--\pageref{lastpage}}
\maketitle

\begin{abstract}
We interpret the correlation between local \sbg positions and ultra-high-energy cosmic ray (UHECR) directions, recently detected by the Pierre Auger Observatory (PAO), in terms of physical parameters: the local density of sources and the magnetic fields governing the UHECR propagation.  We include a Galactic magnetic field model on top of a random extragalactic magnetic field description to determine the level of UHECR deflections expected from an ensemble of source positions. Besides deflections in magnetic fields, we also take into account energy losses with background photon fields as well as spectrum and composition measurements by the PAO. We find consistency between the PAO anisotropy measurement and the local \sbg density for large extragalactic magnetic field strengths with $B > 0.2 \ \rm nG$ (for a coherence length of $1 \ \rm Mpc$) at the $5\sigma$ confidence level. Larger source densities lead to more isotropic background and consequently allow for weaker extragalactic magnetic fields. However, the acceleration of UHECR by such abundant sources is more challenging to motivate. Too large source densities and extragalactic magnetic field strengths, on the other hand, are also disfavored as that decreases the expected level of anisotropy. This leads to upper limits of $B < 22 \ \rm nG$ and $\rho_0 < 8.4 \cdot 10^{-2} \ \rm Mpc^{-3}$ at the 90\% confidence level.

\end{abstract}


\section{Introduction}

The direct test of the origin of ultra-high-energy cosmic rays (UHECRs) by the detection of anisotropies is challenged by the presence of extragalactic magnetic fields (EGMFs) and the Galactic magnetic field (GMF) which lead to deflections of the particles during their propagation. 
While precise EGMF measurements are difficult, the combination of many different observations can provide us with hints for their strength and structure.  The large-scale EGMFs include separate contributions from galaxy clusters, filaments, voids and sheets. Attempts at determining or constraining EGMF strengths and structures can, for example, be found in \citet{1989Natur.341..720K, Barrow:1997mj, Blasi_1999, Kronberg:2007wa, Vallee:2011zz, Ade:2015cva, Pshirkov:2015tua, Vernstrom:2017jvh, OSullivan:2020pll, Vernstrom:2021hru}. This has led to upper limits on intergalactic field strengths in voids of $\lesssim 1$~nG while magnetic fields in the centre of galaxy clusters can get as strong as $\sim10 \mu$G. In addition, simulations of large-scale structure formation can provide model-dependent information about the expected magnetic field strength and structure, see e.g.~\citet{Sigl:2004yk, Dolag:2004kp, Das:2008vb, AlvesBatista:2017vob, Hackstein:2017pex, Garcia:2021cgu} where also estimates for UHECR deflections in these fields are included. The GMF, on the other hand, can be modelled as a combination of a large-scale structured disk component, an extended out-of-plane component and small-scale random components (see e.g.~\citet{Sun:2007mx, Jaffe:2009hh, VanEck:2010ka, Pshirkov:2011um, Jansson:2012pc, Jansson:2012rt, 2017A&A...600A..29T, Boulanger:2018zrk}).

Besides the GMF and EGMF, the deflection of the UHECRs also depends on the rigidity $E/Z$ of the particles, which characterizes the stiffness with respect to magnetic fields; consequently, the UHECR composition affects the propagation.  Since, furthermore, interactions with background radiation fields, such as the cosmic microwave background (CMB), lead to nuclear disintegration and energy losses, potential direct correlations of UHECRs with sources are only expected for a few nearby sources. On the other hand, in the absence of correlations, to avoid significant clustering in the arrival directions of UHECRs at the highest energies, both sufficiently large source densities and extragalactic magnetic field strengths must exist~\citep{Abreu:2013kif}. Such constraints lead to an estimate on the minimum source density of $>2 \cdot 10^{-5} \ {\rm Mpc}^{-3}$, under the assumption that the deflections from the sources are limited to within 30~degrees. Other studies on UHECR arrival directions and their implications have, for example, been performed in \citet{Gorbunov:2002hk, Dermer:2008cy, Takami:2014zva, Eichmann:2017iyr, Capel:2018cnf, Bray:2018ipq, Lang:2020wrr, Harari:2020yml}.

The Pierre Auger Observatory (PAO) has recently detected an excess of UHECRs around the position of nearby \sbgs~\citep{Aab:2018chp,Caccianiga:2019hlc}.\footnote{We note that the objects which the PAO correlated with are \sbgs, utilizing the source class definitions of \citet{Gruppioni_2013}.}  The significance of this anisotropy is at the level of 4.5$\sigma$. 
In the analysis by the PAO an optimal search window (i.e.~a smearing angle around each source using the Fisher distribution) of $\sim 15 \degree$ for a minimal energy threshold of 38~EeV was found for \sbgs; note that the anisotropic fraction of events is significant but small (11\%) compared to the isotropic component. This gives a compelling indication for these galaxies as potential hosts for UHECR acceleration sites.  

A variety of candidate acceleration sites associated with \sbgs have been motivated in the literature. Indeed, particle acceleration up to the ultra-high-energy scale in the termination shock of \sbg winds, when presumed to harbour large magnetic fields, motivate themselves as natural acceleration sites~\citep{Anchordoqui:2018vji}. However, the maximum energy achievable in such a scenario is strongly dependent on the magnetic field strength adopted in the acceleration zone close to the shock; the assumption of more conservative magnetic field strength values reduces the maximum energy scale achievable considerably~\citep{Romero:2018mnb}. Beyond consideration of the galactic wind itself, various other candidate sources associated with \sbgs also exist. Specifically, a higher rate of stellar collapse events than that experienced in normal galaxies should give rise to a higher rate of long gamma-ray bursts (GRBs) in these galaxies. Such GRBs are well-motivated candidate UHECR sources~\citep{Aharonian:2002we}. Additionally, the central active galactic nuclei (AGNs) within these galaxies are also potential candidate sources~\citep{1984RvMP...56..255B}. Note that the PAO has also tested a gamma-ray emitting AGN catalogue for UHECR anisotropies, where a correlation with UHECRs has been found as well -- although it was less significant than for \sbgs~\citep{Aab:2018chp,Caccianiga:2019hlc}. 

With UHECR source candidates ultimately required to possess a source luminosity density level of $4 \cdot 10^{44}~{\rm erg~Mpc}^{-3}~{\rm yr}^{-1}$\citep{1995ApJ...452L...1W,Jiang:2020arb}, a range of source luminosity and density values appear viable, whose product satisfies this criterion. Factoring in the additional requirement on the magnetic luminosity of a candidate source for the acceleration up to energies of 10~EeV, $L_{B}>\beta^{-1}10^{43}~{\rm erg~s}^{-1}$\citep{1976Natur.262..649L}, however, provides additional insights, requiring that candidate sources have a density of $>10^{-6}~{\rm Mpc}^{-3}$. Additionally, the lack of a hard cutoff in the UHECR spectrum~\citep{Taylor:2011ta,Lang:2020wrr} supports the existence of a local UHECR source, consistent with relatively large source-density values.

In this work, we investigate the underlying physical origin of the UHECR anisotropy correlation with \sbgs detected by the PAO. The optimal search window, minimal energy threshold and anisotropic fraction, in combination with the UHECR spectrum and composition measurements, can give an indication for the magnetic field strength between these sources and Earth. For the GMF we use the model from \citet{Jansson:2012pc, Jansson:2012rt}. 
We characterize the local EGMF by a random magnetic field with characteristic strength $B$ and coherence length $\ell_{\mathrm{coh}}$. We interpret the anisotropy result for \sbgs in terms of $B$, $\ell_{\mathrm{coh}}$ and the local source density, which scales the relative contribution of the PAO catalogue and a diffuse (unresolved) contribution needed to describe the large isotropic contribution. 
Compared with \citet{Bray:2018ipq}, instead of only considering the EGMF, we take into account the effects of the GMF and the local source density as well in a completely different analysis method. This allows us not only to obtain an upper limit on the extragalactic magnetic field strength but a lower limit as well. Our methods are similar to the methods used in \citet{Palladino:2019hsk}.

\section{Methods} 

Our goal is to investigate which range of realizations for the magnetic fields in between the nearest \sbgs and Earth can describe the anisotropic signal the PAO observes. After we recapitulate the PAO analysis in Sec.~\ref{sec:auger} we introduce a foreground-background model in Sec.~\ref{sec:bgmodel}, where the foreground is given by the dominant galaxies and the background by an isotropic distribution of the unresolved sources. We use the local density of \sbgs as a scaling parameter, which dictates the relative contribution between the isotropic background and the nearest galaxies; since the total event rate is fixed, a large local density means that the background dominates. We set up simulations of the deflections of UHECRs through the EGMF (Sec.~\ref{sec:EGMF}) and the GMF (Sec.~\ref{sec:GMF}), where we focus on the most dominant sources, investigating the impact of the parameters describing the EGMF. Our results follow a search through a large range of EGMF values and local source densities and are compared with the results from the PAO analysis (Sec.~\ref{sec:augerComp}).

\subsection{Analysis of the Pierre Auger Collaboration}
\label{sec:auger}
For the analysis that found the correlation with \sbgs, the PAO compiled a catalogue of 32 nearby \sbgs~\citep{Aab:2018chp,Caccianiga:2019hlc}. They used the continuum radio emission at 1.4~GHz, produced by $\sim$GeV electrons in these sources, as a proxy for the UHECR flux of these \sbgs. This was done as the gamma-ray luminosity has been shown to correlate almost linearly with the continuum radio luminosity for many \sbgs~\citep{Ackermann:2012vca}. Based on this PAO catalogue they built a probability density map, referred to as their UHECR sky model, which consisted of two components: an isotropic component (a flat probability density map), and an anisotropic contribution from the sources. In this anisotropic component, each source from the PAO catalogue was modelled as a Fisher distribution centred on the source coordinates. The integral of this Fisher distribution for each source was set by the 1.4~GHz radio flux level of each source, with an attenuation factor due to UHECR propagation energy losses being included additionally. The width of this Fisher distribution was determined by the angular width $\theta^{\mathrm{PAO}}$, which was a free parameter and was the same for all sources. The ratio between this anisotropic contribution and the isotropic component was determined by the fraction of all events due to the sources ($f_\mathrm{aniso}^\mathrm{PAO}$), which was the second free parameter of their UHECR sky model. This combined probability density map of isotropic + anisotropic component was then multiplied by the directional exposure of the PAO and the integral of this map was normalized to the number of events ($N_{\mathrm{cr}}=894$ with $E_{\mathrm{th}}^{\mathrm{PAO}}>39$~EeV).

They then performed an unbinned maximum-likelihood analysis, with the likelihood (L) the 'product over the UHECR events of the model density in the UHECR direction'. So
\begin{equation}
	\mathrm{L} = \prod_{n=1}^{N_{\mathrm{cr}}} p_n \, ,
\end{equation}
with $N_{\mathrm{cr}}$ the total number of UHECRs detected by the PAO above $E_{\mathrm{th}}^{\mathrm{PAO}}$ and $p_n$ the probability in the probability density map at the arrival position of particle $n$. The likelihood of the null hypothesis (isotropy) L$_0$ was given by the product of the UHECR events with a flat probability density map multiplied by the directional exposure of the PAO.
The test statistic
\begin{equation}
	\mathrm{TS} = 2\ln \mathrm{L} /\mathrm{L}_0
\end{equation}
for deviation from isotropy gave 'the likelihood ratio test between two nested hypotheses: the UHECR sky model and an isotropic model (null hypothesis)'. This TS was maximized as a function of the two free parameters, $\theta^{\mathrm{PAO}}$ and $f_\mathrm{aniso}^\mathrm{PAO}$. This analysis was repeated for energy thresholds $E_{\mathrm{th}}^{\mathrm{PAO}}$ between 20 and 80~EeV in steps of 1~EeV. The resulting significances were penalized for this scan over threshold energy.

The highest TS, in the case of \sbgs, was found for $E_{\mathrm{th}}^{\mathrm{PAO}}>38$~EeV. The best-fit parameters at this energy threshold were $f_\mathrm{aniso}^\mathrm{PAO}=11^{+5}_{-4}\%$ and $\theta^{\mathrm{PAO}}={15^{\circ}}^{+5^\circ}_{-4^\circ} $. This led to a post-trial significance of $4.5 \sigma$~\citep{Aab:2018chp,Caccianiga:2019hlc}.

\begin{figure*}
	\centering
	\includegraphics[width=0.7\textwidth]{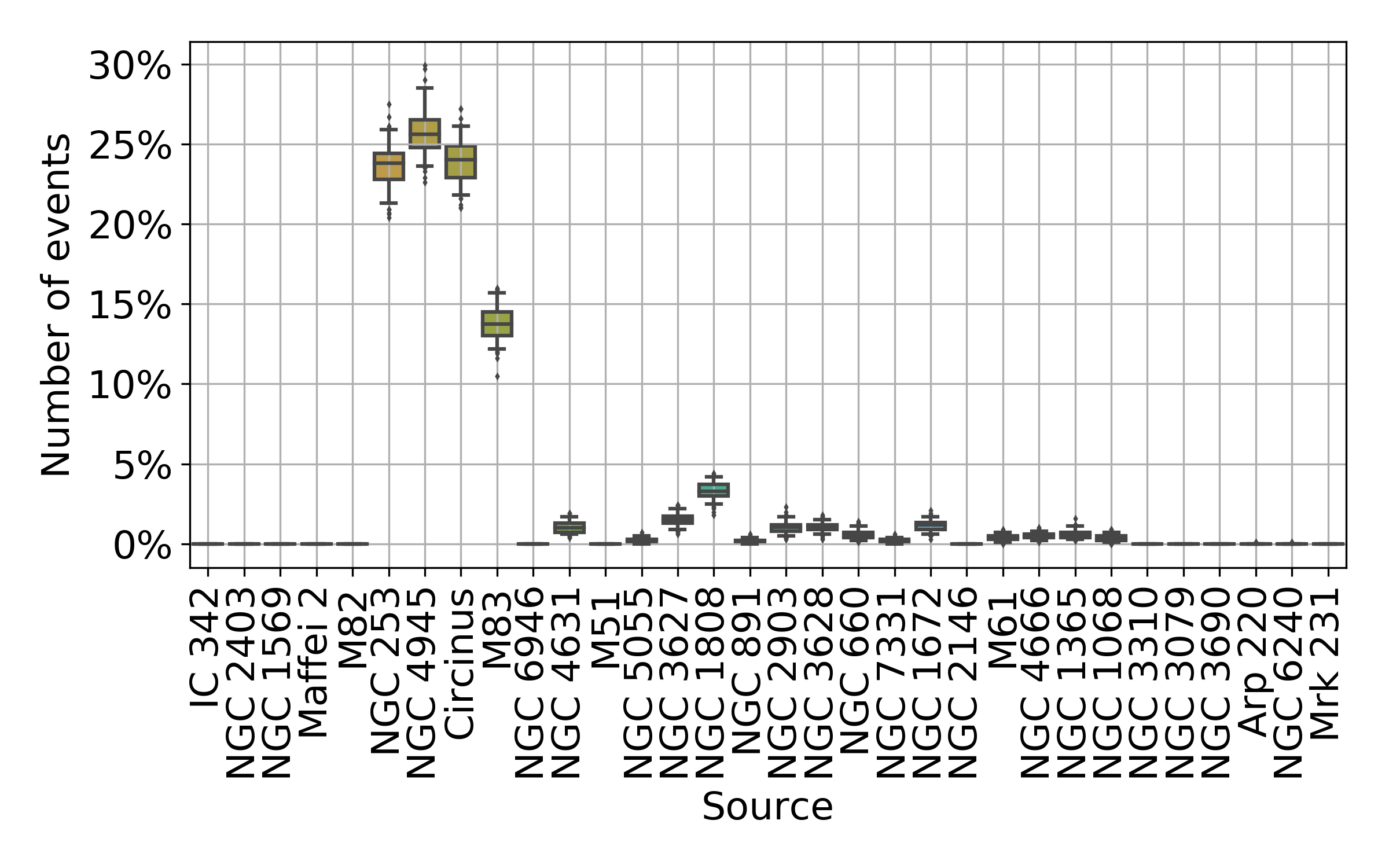}
	\caption{The expected fraction of all catalogue events for each source contained in the PAO catalogue, for 1000 different realizations, represented using a box plot. In this plot, the error bars show the 90\% confidence level, while the boxes show the interval between the 25th and 75th quantile. The sources are ordered from nearest to farthest away. 
In this simulation, the PAO exposure is taken into account, while the cosmic-ray deflections are not considered. Note that NGC~4945, NGC~253, M83 and Circinus account for over 90\% of the signal. The sources without any events are outside the field of view of the PAO.}
	\label{fig:sourcecontr}
\end{figure*}

\subsection{Foreground-background model, and sky map methodology}
\label{sec:bgmodel}

As a first step, assuming only the presence of sources from the PAO catalogue, we compute UHECR sky maps without simulating any deflections, weighting the expected number of events with $1/d^2$ ($d$=distance) and taking into account the exposure of the PAO. In this way, we create sky maps of $N_{\mathrm{cr}}=894$ cosmic rays with energy $E>39$~EeV to be consistent with the analysis of the PAO, as discussed in Sec.~\ref{sec:auger}. In this case, four sources provide $\sim 90\%$ of the signal (see Fig.~\ref{fig:sourcecontr}). These sources are NGC~4945, NGC~253, M83 and Circinus, the four nearest sources in the PAO catalogue that are within the field of view of the PAO. These sources are all within 4.66~Mpc from Earth, while the next source within the field of view of the PAO in the catalogue (NGC~4631) is at a distance of 7.35~Mpc. These four PAO-catalogue sources will, therefore, always, even when adding unresolved sources, be the most promising candidates to produce an anisotropic signal. For the PAO-catalogue sources we, therefore, focus only on these four sources. The relevant properties of these sources are given in Table~\ref{tab:SourceProperties}.

\begin{table}
	\centering
	\caption{Properties of the four most dominant sources, with the sky position in Galactic coordinates.}
	\begin{tabular}{l r r r r r}
  		\hline
     	& NGC~253 & NGC~4945  & Circinus & M83   \\ 
   		\hline
     	Distance $d$ (Mpc) & $3.56$ & $3.72$ & $4.20$ & $4.66$ \\
     	Longitude & $97.364 \degree$ & $305.272 \degree$ & $311.326 \degree$  & $314.584 \degree$ \\
     	Latitude & $-87.965 \degree$ & $13.340 \degree$ & $ -3.808 \degree$  & $31.973 \degree$ \\
  		\hline
	\end{tabular}
  	\label{tab:SourceProperties}
\end{table}

\begin{figure}[t]
	\centering
	\includegraphics[width=0.48\textwidth]{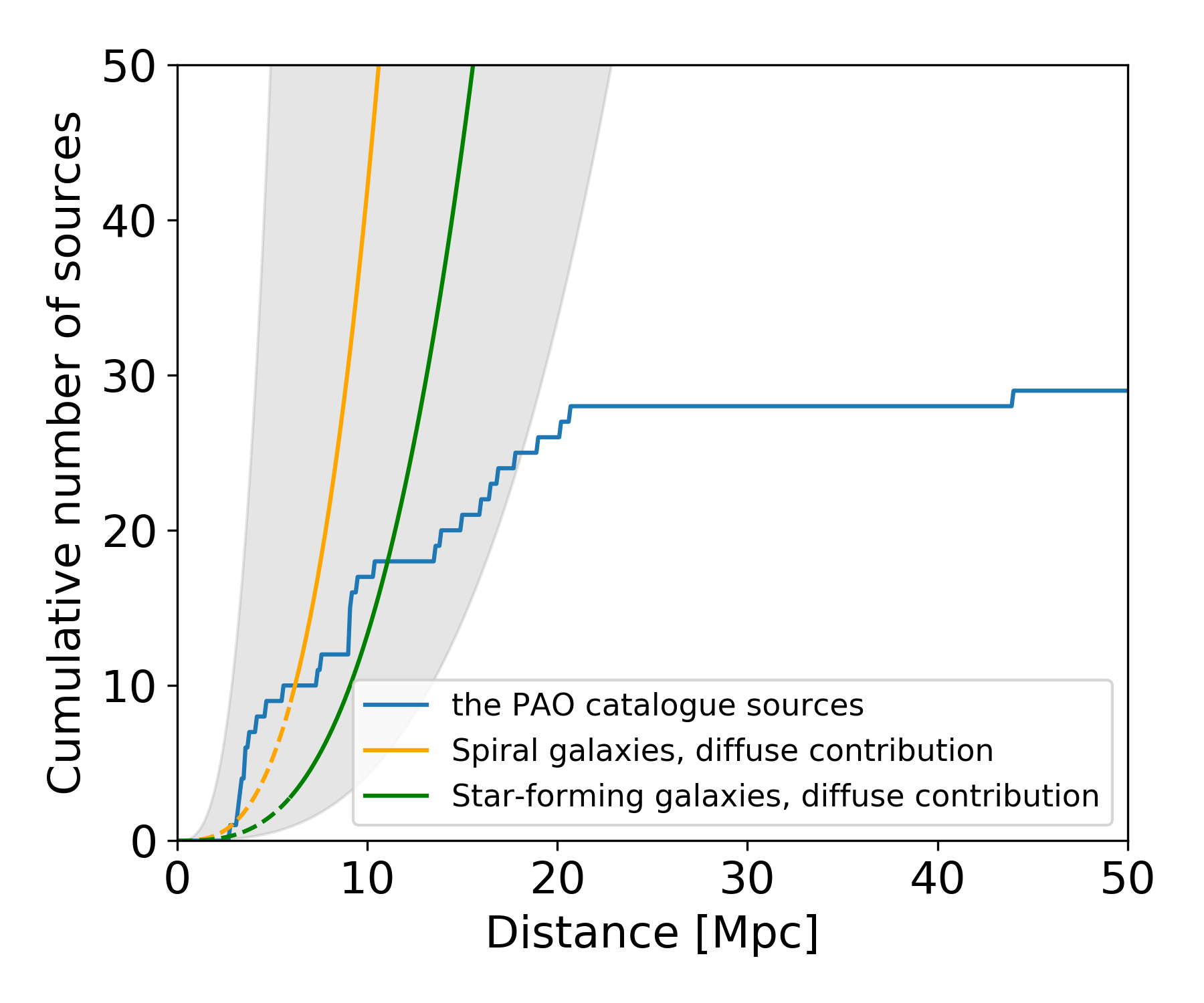}
	\caption{The cumulative number of sources contained in the PAO catalogue, as a function of the distance (blue curve). The orange and green curves represent the extrapolation of the catalogue above 6~Mpc for $\rho_0=10^{-2} \ \rm Mpc^{-3}$ and $\rho_0=3 \cdot 10^{-3} \ \rm Mpc^{-3}$, respectively. These values correspond to the local density of spiral galaxies and \sbgs~\citep{Gruppioni_2013}, respectively. The gray-shaded region corresponds to the entire range of local densities considered in this paper ($\rho_0= 10^{-3} - 10^{-1} \ \rm Mpc^{-3}$).}
	\label{fig:rho0}
\end{figure}

To explain the full UHECR sky, we start from the sources contained in the PAO catalogue (foreground) and we extrapolate the population to large distances, to account for the isotropic background from unresolved sources. The additional sources are placed at random positions in the sky and at distances following a specific value for the local density.
In Fig.~\ref{fig:rho0} we show the cumulative number of PAO-catalogue sources in blue; we note that at small distances low-number fluctuations lead to large deviations from the diffuse flux average. The orange and green curves show possible diffuse flux contributions for $\rho_0=10^{-2} \ \rm Mpc^{-3}$ (orange curve) and $\rho_0=3 \cdot 10^{-3} \ \rm Mpc^{-3}$ (green curve), which are reasonable values for spiral galaxies and \sbgs, respectively~\citep{Gruppioni_2013}. 

The PAO-catalogue sources do not scale in the same way as sources responsible for the diffuse flux beyond about 6-10~Mpc. In fact, the dominant PAO-catalogue sources are all at closer distances than 6~Mpc. We, therefore, combine the cosmic-ray flux from the dominant PAO-catalogue sources with a diffuse contribution for sources beyond a distance of 6~Mpc. For this diffuse contribution we scan over the interval of local densities between $\rho_0= 10^{-3} - 10^{-1} \ \rm Mpc^{-3}$, as  illustrated by the gray-shaded region in Fig.~\ref{fig:rho0}. Let us remark again that the local density scales the unresolved background {\bf only}. Since the total number of events is fixed, a higher local density, therefore, implies a higher contribution from the background compared to the foreground. In the extreme case of a very small local density, we will only have contributions from the PAO-catalogue sources -- and thus a stronger anisotropy. 

We, therefore, create sky maps of $N_{\mathrm{cr}}$ cosmic rays for different realisations of completed source catalogues, sources from the PAO catalogue supplemented with unresolved sources. Next, we introduce the deflection produced by the presence of EMGFs (see Sec.~\ref{sec:EGMF}) and by the presence of both EGMFs and GMFs (see Sec.~\ref{sec:GMF}).

\subsection{Extragalactic propagation of UHECRs}
\label{sec:EGMF}

To determine the expected deflections and energy losses of UHECRs when they travel from their sources to Earth, we simulate their propagation using CRPropa~3~\citep{Batista:2016yrx}. Separate simulations are done, on the one hand, for NGC~4945, NGC~253, M83 and Circinus and, on the other hand, for the isotropically distributed background events. In both cases the energy distribution at the sources is given by
\begingroup
\small
\thinmuskip=\muexpr\thinmuskip*5/8\relax
\medmuskip=\muexpr\medmuskip*5/8\relax
\begin{equation}
	\frac{\text{d}N_i}{\text{d}E} \propto
	\begin{cases}	
		\; f_i  E^{-\gamma} & \text{ for } E < Z_i R_\text{max} \, , \\ 
		\; f_i E^{-\gamma} \exp\left(1  - \frac{E}{Z_i R_\text{max}} \right) & \text{ for } E \geq Z_i R_\text{max} \, , 
	\end{cases}
\end{equation}
\endgroup
with $E$ and $Z_i$ the cosmic-ray energy and charge at the source, respectively. The maximum rigidity at the source is given by $R_\text{max}$, the spectral index by $\gamma$ and the fractions for proton, helium, nitrogen and silicon injection by $f_i$. Table~\ref{tab:CRbestfitparam} gives the values for these parameters, which were obtained by fitting both the spectrum and composition measurements of the PAO~\citep{AlvesBatista:2018zui}. For the Extragalactic Background Light (EBL) the model from \citet{Gilmore:2011ks} is used, and all relevant interactions (photomeson production, photodisintegration, pair production, nuclear decay and adiabatic energy losses) on both the EBL and the CMB are included. 

\begin{table}
\centering
\caption{Best-fit parameters, obtained from \citet{AlvesBatista:2018zui}, used in this work for the UHECR simulations.}
  \begin{tabular}{l c c c c c c c}
  \hline
     $\rho(z)$ & $\gamma$ & $R_\text{max}/\text{V}$ & $f_\text{p}$ & $f_\text{He}$ & $f_\text{N}$ & $f_\text{Si}$ \\ 
   \hline
     SFR & $-1.3$ & $10^{18.2}$ & $0.1628$ & $0.8046$ & $0.0309$ & $0.0018$ \\ 
  \hline
  \end{tabular}
  \label{tab:CRbestfitparam}
\end{table}

For NGC~4945, NGC~253, M83 and Circinus the simulations are done in a 3D setup where the particles are started isotropically at Earth and are detected at spheres with radii equal to their source distances. The deflections in extragalactic magnetic fields of particles from these sources are calculated for specific realizations of random turbulent fields with a Kolmogorov spectrum with RMS magnetic-field strengths $B_\mathrm{RMS}$ and maximum correlation lengths $\ell_{\mathrm{coh}}$. For $B_\mathrm{RMS}$ and $\ell_{\mathrm{coh}}$ we choose ranges appropriate for local extragalactic magnetic fields of $0.1 \ \mathrm{nG} \leq B_\mathrm{RMS} \leq 10 \ \rm nG$ and  $0.2 \ \mathrm{Mpc} \leq \ell_{\mathrm{coh}} \leq 10 \ \rm Mpc$. This range of magnetic-field strengths corresponds to Larmor radii between $\sim 0.5$ and $\sim 50$~Mpc for a typical UHECR rigidity of $R = E/Z = 38/7$~EV. This, therefore, covers the range from small deflections to large deflections for a typical source distance of $\sim 4$~Mpc.

Given that our study here focuses on the UHECR energy scale, the corresponding particle Larmor radii focused on in this work sit close to, or are larger than, the correlation length scale. The results obtained are therefore not anticipated to be strongly sensitive on the actual adopted power-law index of the turbulence spectrum (taken to be Kolmogorov in this work), which controls the power in the turbulence at smaller length scales. We note here that $B_\mathrm{RMS}$ and $\ell_{\mathrm{coh}}$ are degenerate quantities when determining deflections of cosmic rays in magnetic fields. The relevant parameter is $\tilde{B} = B_\mathrm{RMS} \times \sqrt{\ell_{\mathrm{coh}}}$, consistent with predictions from cosmic-ray diffusion in random turbulent magnetic fields~\citep{Neronov:1900zz}. Therefore, we scan over the interval $\tilde{B} = 6 \cdot 10^{-2} - 5 \cdot 10^{1} \ \rm nG \ Mpc^{1/2}$ in logarithmic steps. For each value of $\tilde{B}$, 100 different realizations of the random EGMF are simulated and their results combined. In each realization an equal number of UHECRs is emitted from each source, which are afterwards weighted with $1/d^2$. The source distances and positions in the sky are given in Table~\ref{tab:SourceProperties}. 

Besides deflections in extragalactic magnetic fields, all relevant energy-loss processes are included as well.\footnote{Adiabatic energy losses due to the expansion of the Universe were not included in these 3D simulations, but as this calculation only concerns very local sources and $E>38$~EeV the effect of adiabatic energy losses is negligible.} For each simulated particle registered at one of the detection spheres the deflection angle is determined as the angle between the initial emission direction from the source and the final arrival direction at the detection spheres. This results in a distribution of deflection angles for each value of $\tilde{B}$. 
To compute the sky maps (before GMF deflections), the position of each UHECR originating from one of these four sources is determined by an angle of deflection from the source position  in a random direction following these distributions.

For the UHECRs that form the isotropic background from unresolved sources, the simulations are done in a 1D setup including all relevant interactions and assuming a continuous distribution of identical sources. For these simulations the redshift distribution $\rho(z)$ of \sbgs is given by Star-Formation Rate (SFR) evolution~\citep{Yuksel:2008cu}. 
This gives a distribution of UHECR travel distances for $E>38$~EeV from which specific source realizations can be obtained. The maximum source distance is set to 300~Mpc (ie. $z_{\rm max}\approx 0.07$) as, for $E>38$~EeV, 99.9\% of UHECRs originate from within this distance range. The positions in the sky of these UHECRs are determined randomly, taking into account the exposure of the PAO. 

\begin{figure*}
	\includegraphics[width=0.48\textwidth]{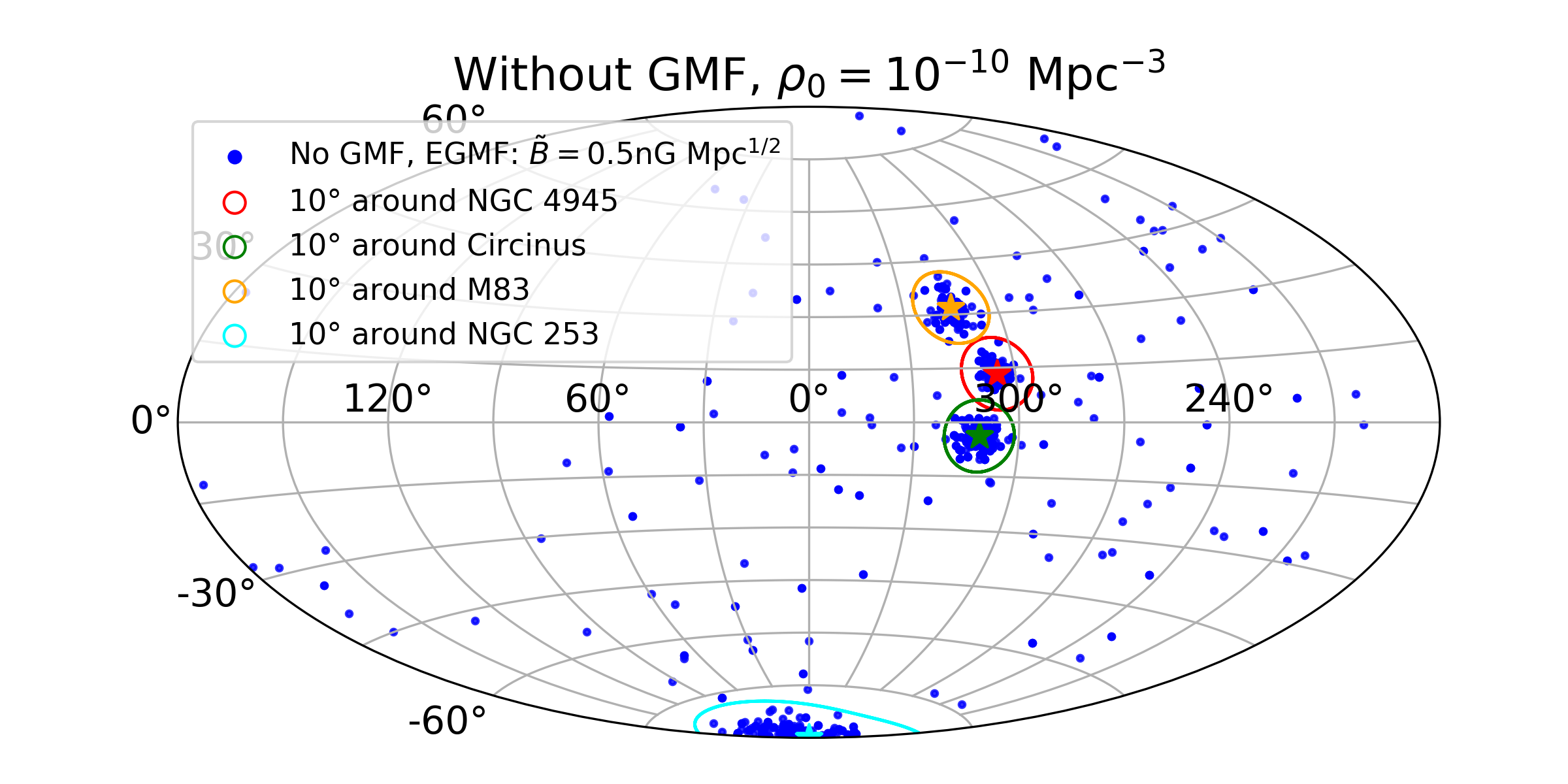} 
	\includegraphics[width=0.48\textwidth]{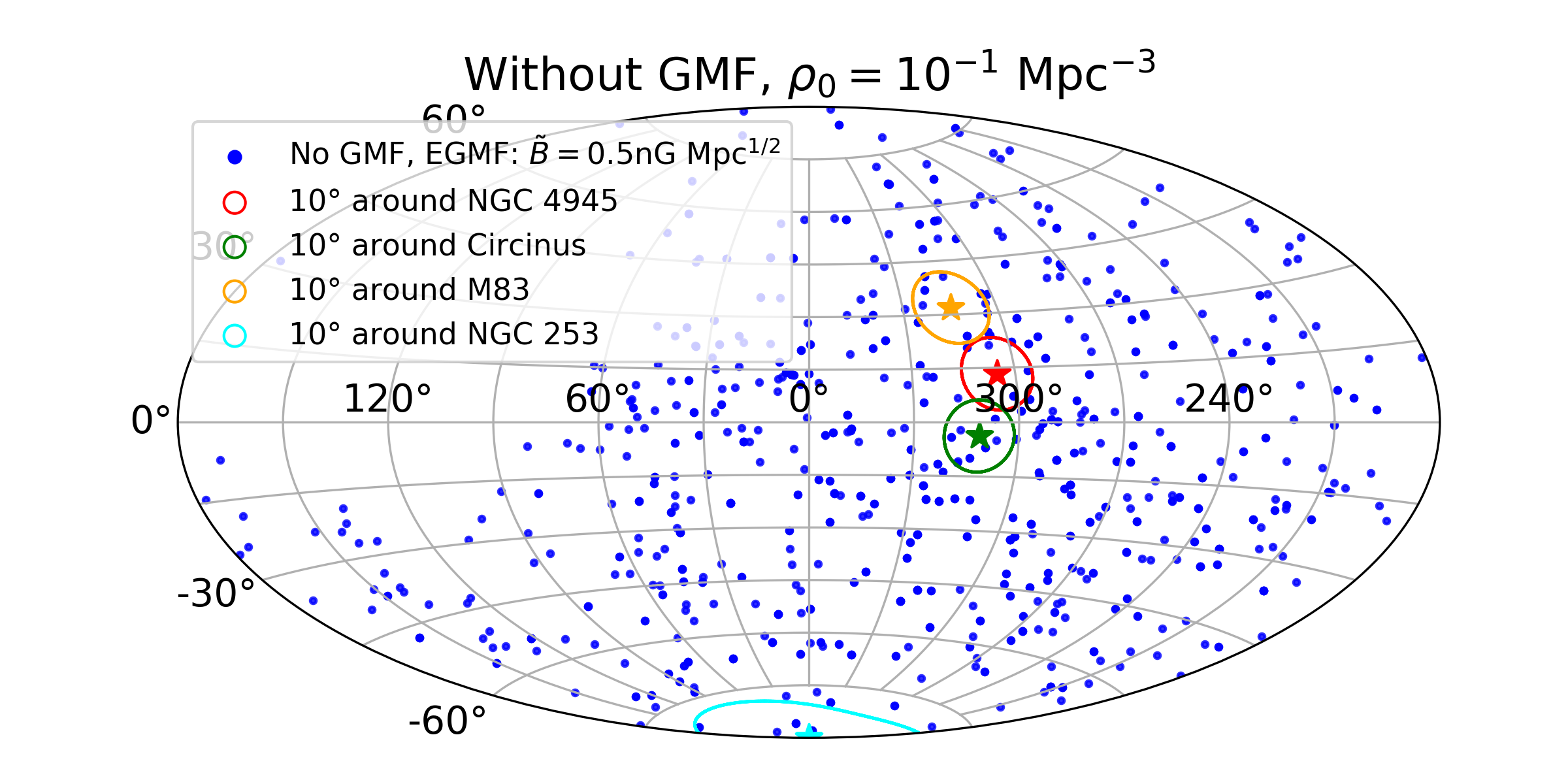}  \\
	\includegraphics[width=0.48\textwidth]{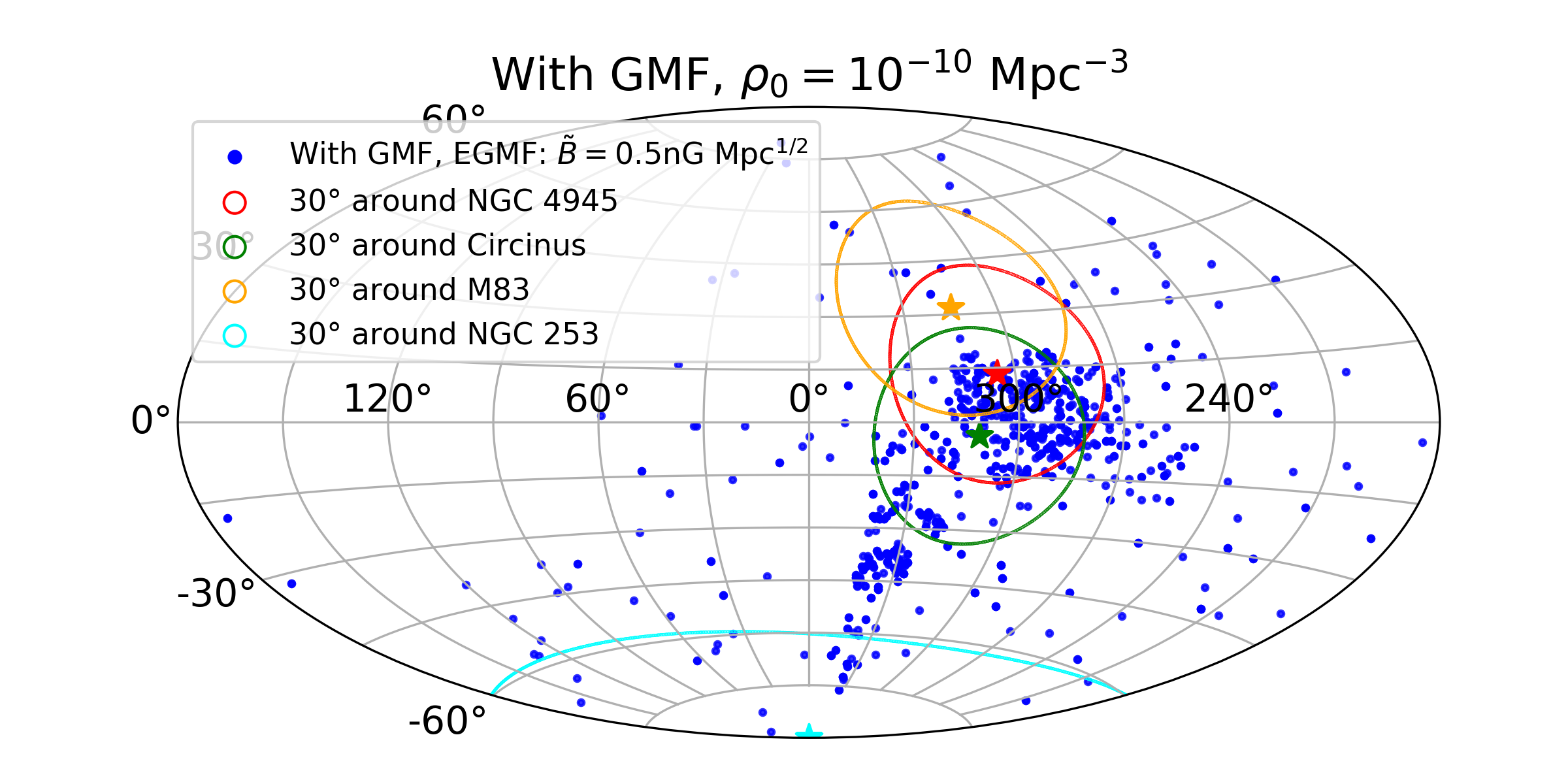}
	\includegraphics[width=0.48\textwidth]{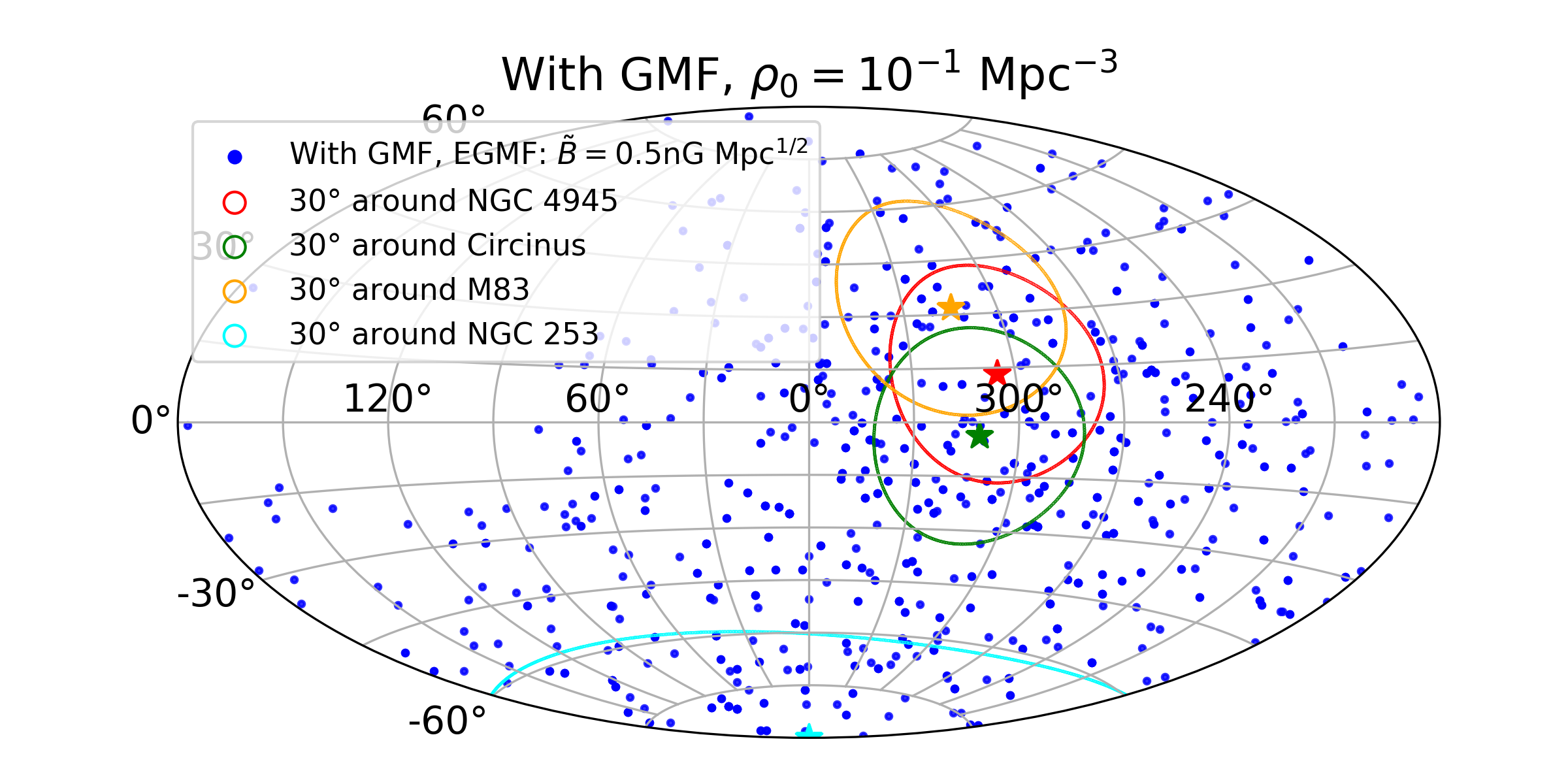}
	\caption{Simulated sky maps, in Galactic coordinates, for an extragalactic magnetic field characterized by $\tilde{B} = 0.5 \ \rm nG \ Mpc^{1/2}$. In the upper panels only the extragalactic magnetic field is present, while in the bottom panels also the Galactic magnetic field is added. The left panels are obtained using only the sources from the PAO catalogue (this can be obtained setting a very low local density). The right panels are obtained using a large value of the local density, corresponding to an isotropic distribution of the sources. In all cases the PAO exposure is taken into account.}
	\label{fig:skyexample}
\end{figure*}

Adding the extragalactic magnetic field, as discussed above, we can simulate sky maps, in which cosmic rays are deflected accounting for their energy, their composition and the source distance. We report in the upper panels of Fig.~\ref{fig:skyexample} two illustrative examples, obtained for $\tilde{B} = 0.5 \ \rm nG \ Mpc^{1/2}$ 
with two extreme/exaggerated values of the local density: a very low value in the left-hand panel (meaning that we essentially only include sources from the PAO catalogue), a very high value in the right-hand panel. In the low local density case, we notice a consistent anisotropy that can be appreciated directly by eye. In the second case the sky map looks much more isotropic. The large empty part of the sky is due to the PAO exposure. 

For each combination of $\tilde{B}$ and $\rho_0$ we produce $10^{4}$ sky maps with $N_{\mathrm{cr}}$ cosmic rays following this method. For these sky maps we determine the angular window (circles of angular radius $\theta_{\rm xs}$ around the positions of the sources) that maximizes the excess of cosmic rays from NGC~4945, NGC~253, M83 and Circinus. Similar to the definition of $f_\mathrm{aniso}^\mathrm{PAO}$ in the analysis by the PAO, this excess over the background is given by the fraction of UHECRs associated to these sources:
\begin{equation}
	f_{\rm xs}(\theta_{\rm xs}) = \frac{N_{\mathrm{xs}}(\theta_{\rm xs}) - b(\theta_{\rm xs})}{N_{\mathrm{cr}}}
\end{equation}
where $N_{\mathrm{xs}}(\theta_{\rm xs})$ is the number of cosmic rays in a certain angular window and $b(\theta_{\rm xs})$ is the expected background in the same angular window. The background is computed assuming an isotropic distribution of sources taking into account the exposure of the PAO. 

\subsection{Comparison with the PAO and scan over $\rho_0$ and $\tilde{B}$}
\label{sec:augerComp}

To be able to compare simulated sky maps obtained with our method with the \sbg correlation results obtained by the PAO we first create $10^5$ sky maps, each with $N_{\mathrm{cr}}=894$ particles, that would give the results for $f_\mathrm{aniso}^\mathrm{PAO}$ and $\theta^{\mathrm{PAO}}$ that the PAO found. These sky maps are produced from probability density maps that were created following the PAO analysis. These probability density maps are created by taking into account the same source PAO catalogue, the same proxy for the UHECR flux, an attenuation factor due to energy losses derived from UHECR propagation simulations and the directional exposure of the PAO in the same way as described in Sec.~\ref{sec:auger}. The search radius and anisotropic fraction, needed for the creation of the probability density maps, are fixed to the best-fit parameters that the PAO found ($f_\mathrm{aniso}^\mathrm{PAO}=11^{+5}_{-4}\%$ and $\theta^{\mathrm{PAO}}={15^{\circ}}^{+5^\circ}_{-4^\circ} $). On the $10^5$ sky maps created in this way, we run our analysis and determine the angular window that maximizes the excess fraction of cosmic rays. In this way, we find a maximum excess of $f_{\rm xs}^{\rm PAO} \pm \delta f_{\rm xs}^{\rm PAO} = 0.050 \pm 0.020$ using angular windows with radii of $\theta_{\rm xs}^{\rm PAO} \pm \delta \theta_{\rm xs}^{\rm PAO} = 31^\circ \pm 11^\circ$. 
These are, therefore, the values for the results of our analysis that correspond to the best-fit parameters of the PAO analysis. For each simulated sky map we produce with our method we determine the angular window and maximum excess of cosmic rays and compare these values with the values that correspond to the best-fit parameters of the analysis by the PAO. The differences between $f_\mathrm{aniso}^\mathrm{PAO}$ and $\theta^{\mathrm{PAO}}$ on the one hand and $f_{\rm xs}^{\rm PAO}$ and $\theta_{\rm xs}^{\rm PAO}$ on the other hand are due to the differences between the PAO analysis and our analysis, mainly caused by the different definition of the angular windows $\theta$.
  
\begin{figure}
	\centering
	\includegraphics[width=0.45\textwidth]{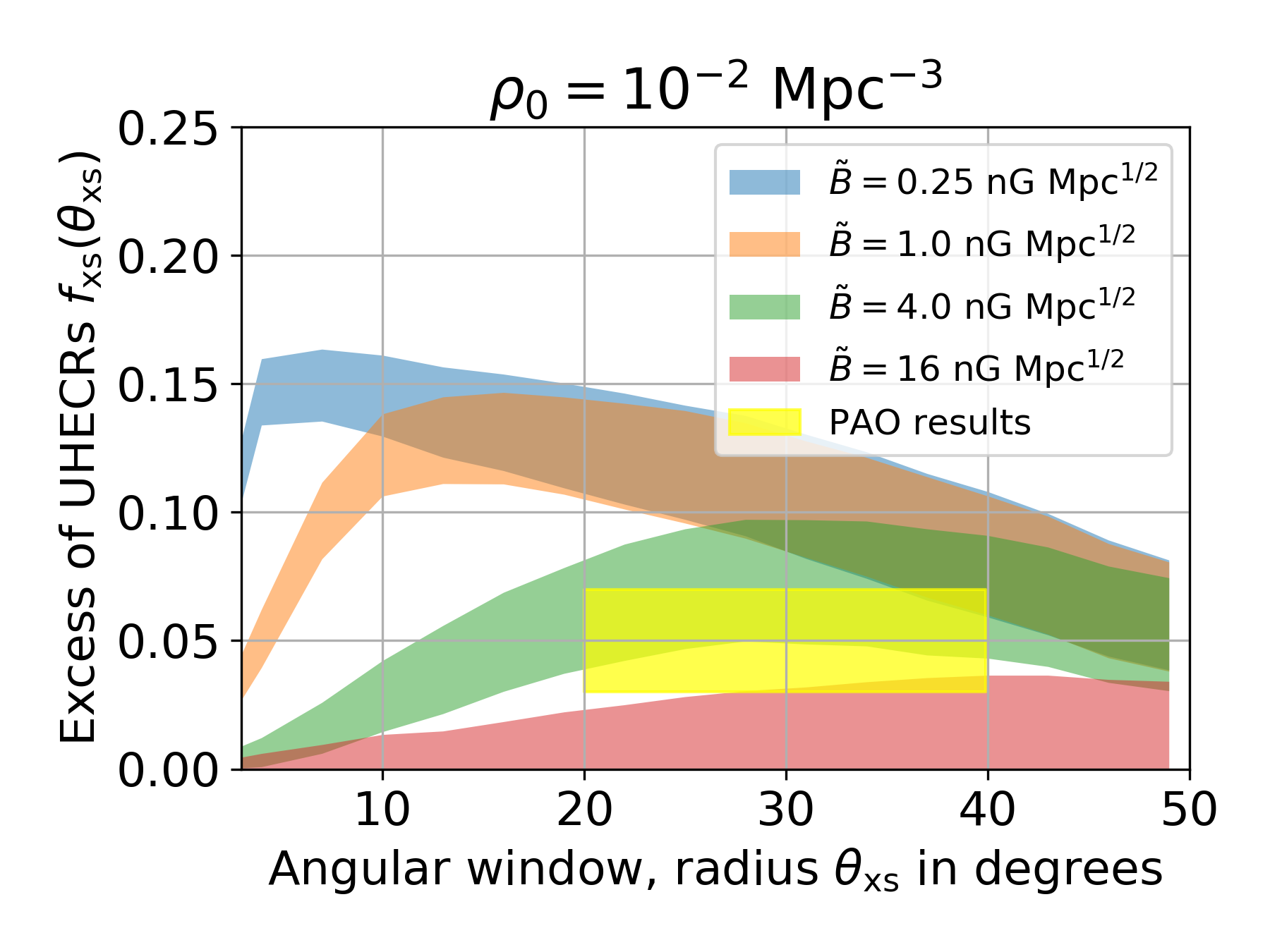}
	\caption{Excess number of cosmic rays over the background in the angular window (for radii of this angular window between $3^{\circ}$ and $50^{\circ}$), as a function of the extragalactic magnetic field parameter $\tilde{B}$. In this example the Galactic magnetic field is not included. The local density is fixed to $\rho_0=10^{-2} \ \rm Mpc^{-3}$. The thickness of the colored regions gives the $1\sigma$ upper bound and lower bound of the 1000 different realizations of sky maps simulated for each combination of $\tilde{B}$ and $\rho_0$. The results corresponding to the analysis performed by the PAO in terms of our parameters are indicated as well (yellow region).}
	\label{fig:angwind}
\end{figure}

In Fig.~\ref{fig:angwind} we show an example for how this comparison is done. Here we fix the local source density to $\rho_0=10^{-2} \ \rm Mpc^{-3}$ and we vary the magnetic-field parameter $\tilde{B}$, showing four different configurations. For each configuration we illustrate the excess fraction of cosmic rays over the background ($f_{\rm xs}(\theta_{\rm xs})$) around NGC~4945, NGC~253, M83 and Circinus, as a function of the angular window. The yellow box shows the results obtained from the analysis by the PAO; therefore, for this specific scenario, a strong extragalactic magnetic field is required to be in agreement with the observations (neglecting deflections in the GMF).

We next explore the landscape of this source anisotropy level through a scan over the local source density $\rho_0$ and effective EGMF values $\tilde{B}$. We compute the maximum excess fraction of cosmic rays $f_{\rm xs}^{\rm max}$ and the corresponding angular window $\theta_{\rm xs}^{\rm max}$ for each combination of $\rho_0$ and $\tilde{B}$. We then determine the $\chi^2$ for this parameter set as follows:
\begin{equation}
		\chi^2(\tilde{B},\rho_0) = \frac{[\theta_{\rm xs}^{\rm PAO}-\theta_{\rm xs}^{\rm max}(\tilde{B},\rho_0)]^2}{(\delta \theta_{\rm xs}^{\rm PAO})^2}
		+ \frac{[f_{\rm xs}^{\rm PAO} - f_{\rm xs}^{\rm max}(\tilde{B},\rho_0)]^2}{(\delta f_{\rm xs}^{\rm PAO})^2}
\end{equation}
Finally, we convert the $\chi^2$ for 2 degrees of freedom into the corresponding Gaussian number of sigmas.
We scan the local source density over the range $\rho_0= 10^{-3} - 10^{-1} \ \rm Mpc^{-3}$ and the EGMF parameter in the range $\tilde{B} = 6 \cdot 10^{-2} - 5 \cdot 10^{1} \ \rm nG \ Mpc^{1/2}$.

In our analysis, the excess of UHECRs in the correlation study of the PAO that we are comparing our results with is given by $f_{\rm xs}^{\rm PAO} \pm \delta f_{\rm xs}^{\rm PAO} = 0.050 \pm 0.020$. This is compatible with $f_{\rm xs} = 0$ (complete isotropy) within $\sim 2.5\sigma$. It is, therefore, with our analysis and the current results of the PAO, not possible to obtain upper bounds on $\tilde{B}$ and $\rho_0$ at a larger confidence level (C.L.) than that. For lower bounds, on the other hand, this problem does not occur.

\subsection{Deflections in the Galactic magnetic field}
\label{sec:GMF}

Beside the EGMF, UHECRs can be deflected in the Galactic magnetic field as well. To simulate the GMF deflections the GMF lensing technique of CRPropa~3 is used~\citep{Bretz:2013oka}. For this method lenses are created by backtracking cosmic rays through a GMF model from Earth to the edge of our Galaxy for the whole range of relevant rigidities and isotropic emission directions from Earth. From these simulations matrices (lenses) are created for each rigidity that convert the arrival direction at the edge of the Galaxy to possible arrival directions at Earth.

The GMF model we use is the full JF12 model~\citep{Jansson:2012pc, Jansson:2012rt}. This GMF model consists of a regular large-scale field and a component that describes random fields. The regular field consists of a disk field and a halo component build up out of a toroidal part and an x-shaped field, while the random field consists of randomly-oriented small-scale fields as well as striated random fields whose orientation is aligned over a large scale. 

To apply the GMF lens for this GMF model, first an UHECRs sky map is created from the extragalactic 3D simulations by determining the deflection angle of each simulated particle. Then, each particle is placed on the sky map at its deflection angle away from its original source position in a random direction. This gives an UHECR sky map at the edge of our Galaxy for which the energy and charge information of the arriving particles is still available, which is necessary to determine the GMF deflections. Then the lensing technique is run on this sky map, resulting in a probability density map for UHECR arrival directions at Earth. This is performed for each combination of $\tilde{B}$ and $\rho_0$ so that probability density maps are obtained for each EGMF scenario. In the bottom panels of Fig.~\ref{fig:skyexample} we show two illustrative examples including deflections in the GMF. In both cases the EGMF parameter $\tilde{B}$ and the local density $\rho_0$ are fixed to the same values as the sky maps in the upper panels (without GMF deflections).

\section{Results}

\begin{figure*}[t]
	\includegraphics[width=0.32\textwidth]{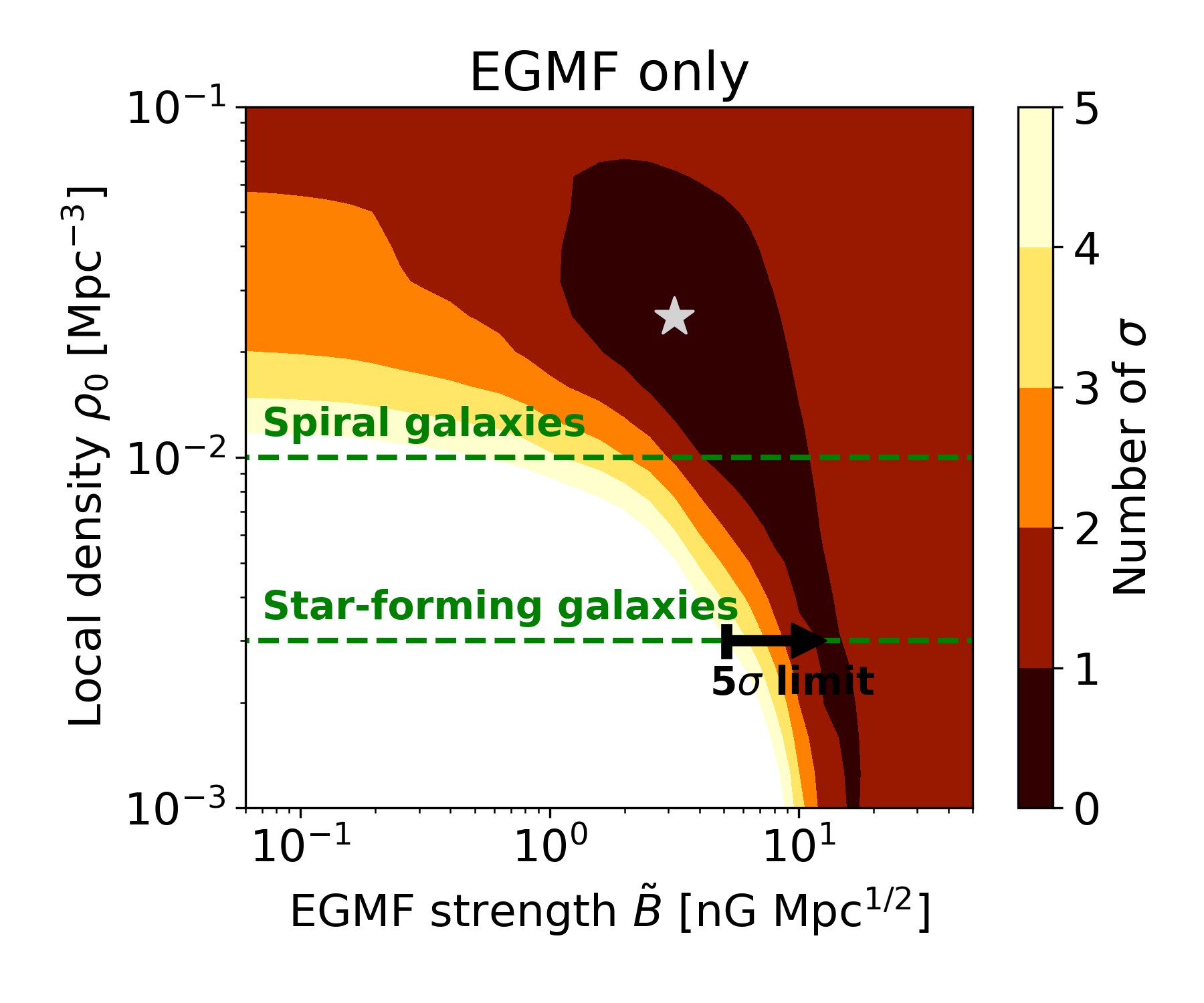} 
	\includegraphics[width=0.32\textwidth]{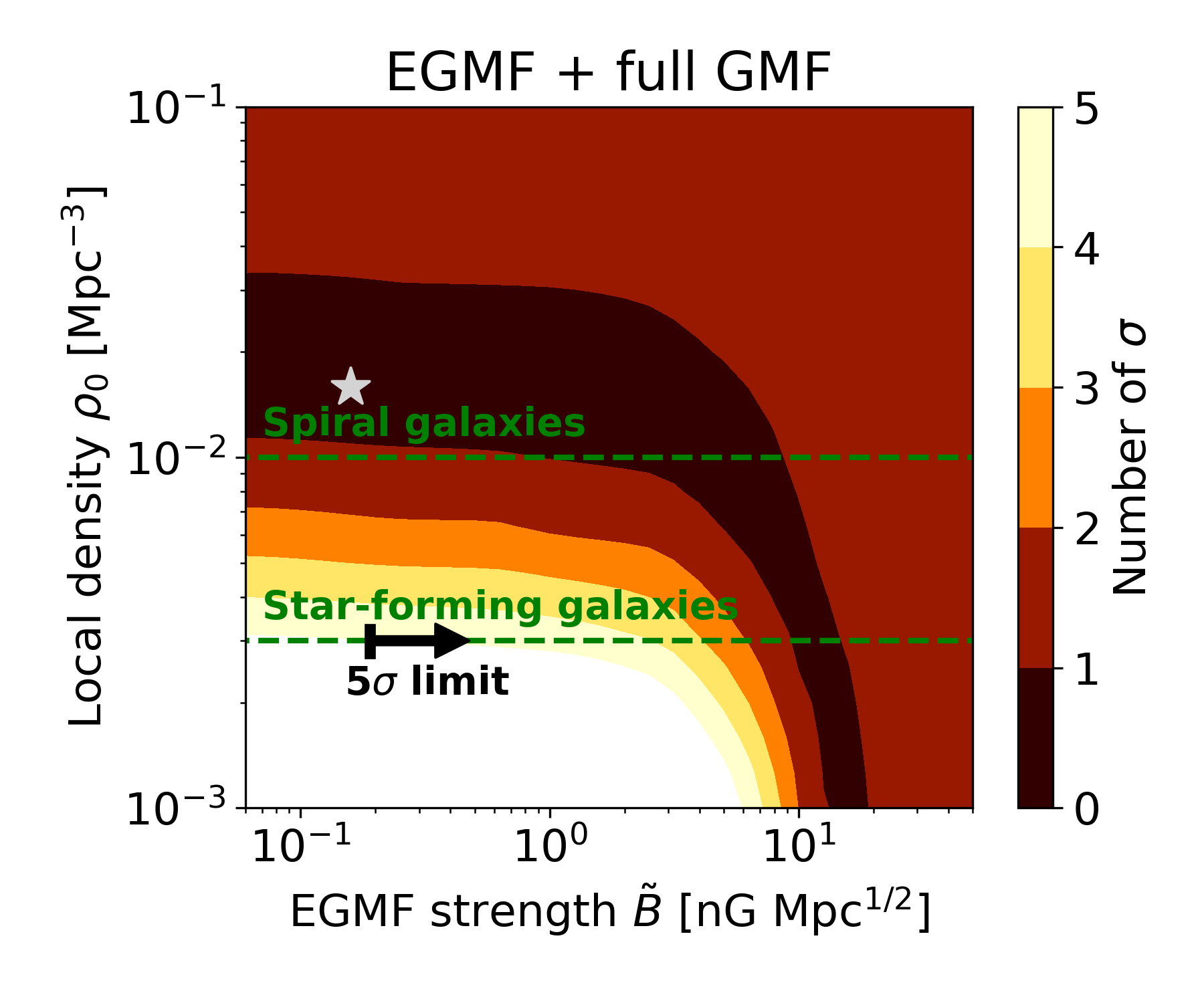}
	\includegraphics[width=0.32\textwidth]{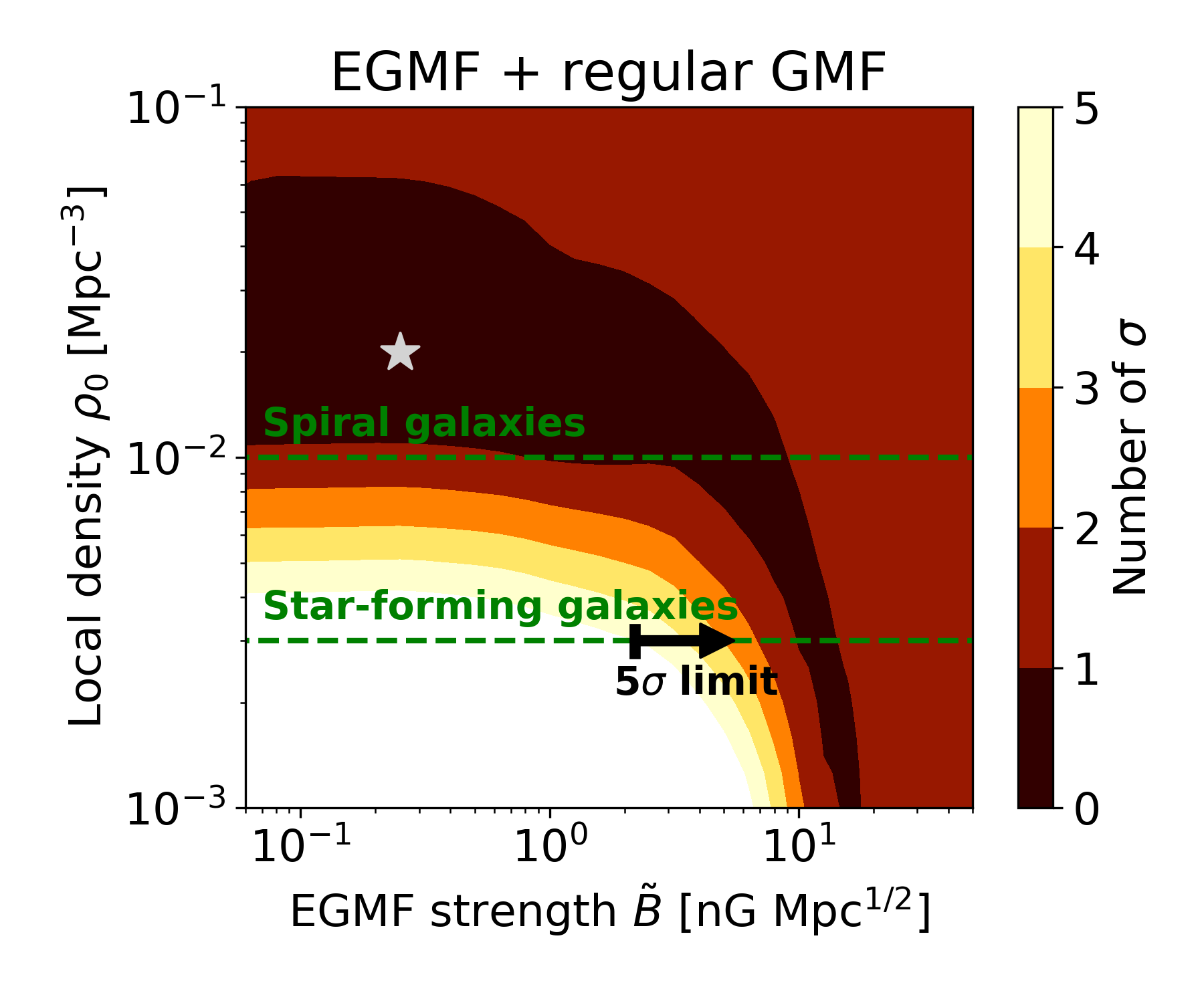}
	\caption{Allowed region in EGMF strength and local density for three scenarios: EGMF only (left panel), EGMF and the full GMF (middle panel), and EMGF + only the regular GMF (right panel), where the GMF is fixed within each panel. 
	By comparing our results with the results obtained by the PAO, we represent the regions that are $N\sigma$ away from the PAO results. The number of sigma is obtained using the square root of $\chi^2$ for two degrees of freedom. The best-fit point of each specific scenario is indicated by the star. We also denote the local densities corresponding to spiral galaxies and to \sbgs of a certain mass~\citep{Gruppioni_2013}, including the 5$\sigma$ lower limit on $\tilde{B}$ for this source density for each specific scenario.}
	\label{fig:main_comp}
\end{figure*}

From the illustrative examples in Fig.~\ref{fig:skyexample} the main effects of the source density and the Galactic magnetic field are immediately apparent. For the low local density case without GMF deflections (top-left panel), the anisotropy is visibly appreciated, with most of the cosmic rays accumulating around the four sources: NGC~4945, Circinus, M83, and NGC~253, within a radius of $10^\circ$. In contrast, for the corresponding high local density case (top-right panel), the anisotropy signal appears too dilute to be easily observable. The situation becomes very different when the GMF model is included in addition (bottom panels). For the low local density case, we notice that now the deflection is much larger. Moreover, we also observe a systematic shift in the position of cosmic rays produced by the two sources located furthest from the Galactic plane, namely NGC~253 and M83. The association with these sources is therefore to some extent lost, since cosmic rays are no longer distributed symmetrically around their source positions, with the centres of the arrival-direction distributions instead shifted in the direction of the Galactic plane.

The results of full scans over $\tilde{B}$ and $\rho_0$ for three different scenarios are shown in Fig.~\ref{fig:main_comp}. 
The level of agreement obtained from a comparison of the model-predicted correlation with that actually measured by the PAO (see Sec.~\ref{sec:augerComp}) are shown. In the same figures we also indicate the typical local densities expected for spiral galaxies and \sbgs~\citep{Gruppioni_2013}. The limits and best-fit points that are obtained for the three different scenarios are given in Table~\ref{tab:results}.

\begin{table*}
\centering
\caption{Limits on $\tilde{B}$ and $\rho_0$ and best-fit point for the three scenarios shown in Fig.~\ref{fig:main_comp}.}
  \begin{tabular}{m{2.7cm} m{4.5cm} m{4.5cm} m{4.5cm}}
  \hline
      & EGMF only & EGMF + full GMF & EGMF + regular GMF \\ 
   \hline
     $5\sigma$ lower limit on $\tilde{B}$ for $\rho_0 = 3 \cdot 10^{-3} \ \rm Mpc^{-3}$ & $\tilde{B} > 5.1 \ \rm nG \ Mpc^{1/2}$ & $\tilde{B} > 0.19 \ \rm nG \ Mpc^{1/2}$ & $\tilde{B} > 2.2 \ \rm nG \ Mpc^{1/2}$ \\ 
     \hline
     Best-fit point & $\tilde{B}=3.2\ \rm nG \ Mpc^{1/2}$; \newline $\rho_0=2.5 \cdot 10^{-2} \ \rm Mpc^{-3}$ & $\tilde{B}=0.16 \ \rm nG \ Mpc^{1/2}$; \newline $\rho_0=1.6 \cdot 10^{-2} \ \rm Mpc^{-3}$ & $\tilde{B}=0.25\ \rm nG \ Mpc^{1/2}$; \newline $\rho_0=2.0 \cdot 10^{-2} \ \rm Mpc^{-3}$ \\ 
     \hline
     90\%~C.L.~region & $\tilde{B} < 22 \ \rm nG \ Mpc^{1/2}$; \newline $\rho_0 < 8.4 \cdot 10^{-2} \ \rm Mpc^{-3}$ & $\tilde{B} < 22\ \rm nG \ Mpc^{1/2}$; \newline $\rho_0 < 4.3 \cdot 10^{-2} \ \rm Mpc^{-3}$ & $\tilde{B} < 21\ \rm nG \ Mpc^{1/2}$; \newline $\rho_0 < 7.1 \cdot 10^{-2} \ \rm Mpc^{-3}$ \\
  \hline
  \end{tabular}
  \label{tab:results}
\end{table*}

In the left-hand panel of Fig.~\ref{fig:main_comp} we report the results obtained adopting only an extragalactic magnetic field (ignoring any Galactic magnetic field effects). In the middle panel the results are reported for deflections in both the EGMF and the GMF (full JF12 model~\citep{Jansson:2012pc, Jansson:2012rt}). The GMF produces a strong coherent deflection of the cosmic rays away from their source positions, requiring a reduced level of deflection by the extragalactic magnetic field, compared to the EGMF only case, in order to be in agreement with the PAO results. The results in the right-hand panel differ from the results obtained for the EMGF + GMF scenario shown in the middle panel, in that the GMF model employed now only contains the regular components of the JF12 model, with the random components of the GMF model being left out. The effect of the regular GMF components is to shift the arrival directions away from the source position in specific directions, while the random components 'spread out' the deflections in all directions.

Fixing the local density to the local density of \sbgs\footnote{The local density of \sbgs, $\rho_0 = 3 \cdot 10^{-3} \ \rm Mpc^{-3}$,  is obtained from~\citet{Gruppioni_2013}. However, if another survey finds a different source density for \sbgs, the corresponding $5\sigma$ limit on $\tilde{B}$ can still be obtained from Fig.~\ref{fig:main_comp}.} we find a $5\sigma$ lower limit on the local EGMF, for one degree of freedom, for all three scenarios (see Table~\ref{tab:results}).
In this case, the scenario with the complete GMF model provides the most conservative lower limit ($\tilde{B} > 0.19 \ \rm nG \ Mpc^{1/2}$) as there the cosmic-ray deflections are the largest. Additionally, considering the full range for $\rho_0$, the best-fit points are indicated for all three scenarios. In all three cases we find a local density close to or even larger than that of spiral galaxies. Regarding only the 90\% confidence level region, upper bounds on both the EGMF strength and the local source density are obtained again for all three scenarios. The most conservative upper bounds are from the scenario without a GMF ($\tilde{B} < 22 \ \rm nG \ Mpc^{1/2}$ and $\rho_0 < 8.4 \cdot 10^{-2} \ \rm Mpc^{-3}$) as there the total deflections are the smallest. 

Furthermore, in the EGMF only scenario, an anti-correlation between the local density and the EGMF strength is observed. Such a relation is expected since a larger local-density value increases the number of contributing sources, which reduces the anisotropy level. Likewise, weaker EGMFs lead to smaller deflections, which increases the anisotropy level. A similar anisotropy level is, therefore, maintained through the source density and EGMF strength changing in a reciprocal manner. In the scenarios with GMF deflections, the anti-correlation between $\tilde{B}$ and $\rho_0$ is less visible since, for large values of $\rho_0$, the GMF alone already deflects the UHECRs sufficiently to agree with the PAO results. 

Although our results are applicable for the UHECR spectra and composition scenario provided in Table~\ref{tab:SourceProperties}, consideration of the consequences of a departure of this assumption provides insight. As was found in \citet{Taylor:2015rla}, a negative evolution scenario (or a local source overdensity such as in \citet{2002AJ....124..675C}, which effectively operates in this way) allows considerably softer source spectra. A scenario in which local $<20$~Mpc sources with softer spectra than those considered here dominate the UHECRs for $E>38$~EeV, could therefore also provide agreement with the PAO results. Since such a case would require even stronger local magnetic fields than those found here in order to ensure sufficient cosmic-ray isotropisation from such local sources, the lower bound on the EGMF that we have found can be considered conservative.

In App.~\ref{app:protons} we also show the results for a different composition of UHECRs, namely a pure-proton case. While this scenario is already excluded by the composition measurements of the PAO, it gives an indication for how the results depend on the composition, and for how the results change if the deflections are minimized. In this case the lower limit on the local EGMF for the local source density of \sbgs becomes much stronger. A large value for the local density is required, even for strong EGMF values.

\section{Summary and discussion}

The Pierre Auger Collaboration has detected an excess of UHECRs around the position of nearby \sbgs, with a significance of 4.5$\sigma$.  
In this paper we have interpreted these results theoretically in terms of what can be learned about the extragalactic magnetic fields (EGMFs) -- whilst accounting for the potential effects of the Galactic magnetic field (GMF).

Since the detected anisotropic ``foreground'' only consists of a small fraction of UHECR events (11\%), we have introduced an isotropic background contribution to take into account the unresolved sources using the local source density as a free parameter. Note that since the total number of events is fixed, a higher local density increases the isotropic background fraction. We have focused on the fact that the anisotropic foreground is dominated by four sources: NGC~4945, NGC~253, M83 and Circinus, which collectively account for most of the anisotropic signal.
We have sampled cosmic rays from the combined foreground-background source maps, taking into account UHECR interactions during their propagation and the position in the sky in order to account for the PAO exposure. We have subsequently determined the expected deflection for all UHECRs due to the EGMF and GMF using source-emission spectra and mass compositions that fit measurements on the UHECR energy spectrum and mass composition at Earth.
While for the GMF we have used the JF12 model, we have assumed that the EGMF is described by random turbulent fields following a Kolmogorov spectrum. 
Consequently, the (effective) extragalactic magnetic field $\tilde B = B_{\mathrm{RMS}} \times \sqrt{\ell_\mathrm{coh}}$, degenerate with the coherence length, and the local source density $\rho_0$ are our main theoretical parameters to interpret the correlations observed by the PAO. 

Due to the relatively small distance to the dominant sources ($< 5$~Mpc), no large-scale structure in the EGMF is expected in between the sources and the Milky Way. The EGMF in between the dominant sources and the Milky Way is, therefore, well described by random turbulent fields. Whether a specific structured EGMF model agrees with the UHECR correlations with \sbgs as observed by the PAO can, therefore, be determined from Fig.~\ref{fig:main_comp}. In this case, the average EGMF strength and correlation length in between the dominant sources and the Milky Way in the specific structured EGMF model would provide a good estimate for $\tilde{B}$. The discussion of specific EGMF models in detail is outside the scope of this work.

Assuming that the background UHECRs indeed come from the same class as the PAO catalogue sources, a logical choice for the local source density value is that for \sbgs.
For this source density we find a solid lower limit on the EGMF $\tilde{B} > 0.19 \ \rm nG \ Mpc^{1/2}$ (at the $5\sigma$ confidence level) independent of the specific EGMF+GMF scenario considered.  
The reason is that \sbgs\ are comparatively rare (i.e., have a small source density), requiring the relatively isotropic distribution observed by the PAO has been generated by deflections in magnetic fields.
Current lower limits for intergalactic void fields in general, from the non-detection of electromagnetic cascades initiated by gamma rays, are set at $B > 10^{-8}$~nG~\citep{Neronov:1900zz, Tavecchio:2010ja,Taylor:2011bn}. 
Our lower limit, therefore, severely restricts the allowed parameter space for the EGMF strength.
However, our limit is only applicable for local extragalactic magnetic fields between the four dominant nearby sources and the Milky Way. The regions surrounding these galaxies can be dominated (or at least heavily contaminated) by magnetised haloes produced by star-formation winds, if these winds are magnetised, or by previous activity of the galactic nucleus (see e.g.~\citet{Marinacci:2015dja, Vazza:2017qge,  Garaldi:2020xos, Attia:2021ywb}). In that case, the limits obtained in this work limit the magnetisation from galaxies within their halo, rather than volume filling primordial fields. Furthermore, note that the relatively strong EGMF strength indicated by the lower limit is consistent with what is required to explain the UHECR dipole measured by the PAO~\citep{AugerDipole} at lower energies~\citep{Lang:2020qhh}.

If the constraint on the local source density is released, an anti-correlation between source density and EGMF field emerges: isotropization can either be generated by a strong EGMF or large local densities.  Since a too strong isotropization destroys the observed correlation with the foreground sources, a more general upper limit on both the EGMF and the local source density can be placed at the 90\% C.L., which are $\tilde{B} < 22 \ \rm nG \ Mpc^{1/2}$ and $\rho_0 < 8.4 \cdot 10^{-2} \ \rm Mpc^{-3}$.
These limits correspond to our most conservative case, which means that the deflections only come from the EGMF. Stronger bounds are found if the GMF is included, where the dominant effect comes from the halo component of the GMF. Since this component is rather uncertain and systematic shifts of two of the sources off the Galactic plane cannot be captured well by our/the PAO approach, we consider the EGMF only case most conservative here.
The obtained limit on the extragalactic magnetic field is somewhat larger than the limit given by the Planck Collaboration, from the CMB temperature and polarization spectra, which is at the nanoGauss level~\citep{Ade:2015cva}. 

We also find that the best-fit point for the two-parameter fit is for a source density close to, or even denser than, that of spiral galaxies. Considering specifically the four most significant sources driving the PAO anisotropy signal; NGC~253, M83, Circinus and NGC~4945 are all local spiral galaxies in the ``{\it Council of Giants}''. While Circinus and NGC~4945 are both Seyfert galaxies, a class of mostly spiral galaxies that have compact bright cores, NGC~253 and M83 are intermediate spiral galaxies (with high central star-formation rates). Only a few spiral galaxies closer than NGC~253 are not included in the PAO catalogue. Therefore, the fact that such regular type galaxies drive the correlation is fully compatible with these large source density findings, which push up the isotropic background fraction from more distant sources. However, this would imply that, barring selection effects or statistics, all (or at least most) spiral galaxies contain UHECR accelerators, which is hard to motivate from the current knowledge of UHECR acceleration. In fact, the Milky Way itself is also a spiral galaxy but no indications exist that the Milky Way currently contains an UHECR source~\citep{Giacinti:2011ww,Abreu:2012ybu}. In addition, it is interesting that both NGC~4945 and Circinus have recently been motivated as cosmic-ray accelerators contributing to the astrophysical neutrino flux~\citep{Kheirandish:2021wkm}.

In conclusion, assuming that the UHECR correlations with local \sbgs that the PAO has found are real, two different scenarios seem possible. Either \sbgs are indeed the sources of UHECRs, in which case rather strong extragalactic magnetic fields between these sources and the Milky Way need to be present, or the UHECRs are produced by sources more numerous than \sbgs ($\rho_0 > 3 \cdot 10^{-3} \ \rm Mpc^{-3}$), which is however rather challenging to motivate from the current knowledge of possible UHECR acceleration sites. 

With increased statistics, the significance of the PAO's result is expected to soon reach 5$\sigma$, within the timescale of a few years. Therefore, it will be possible in the future to have a refined version of this analysis, also including other Galactic magnetic field models, since in this paper only one benchmark model for the GMF has been considered. On the other hand, we find that the GMF leads to systematic offsets of the source positions which cannot be captured by the Fisher template used in the PAO analysis; therefore, future analyses of UHECR anisotropies could benefit from considering Galactic deflections more directly.

\section*{Acknowledgements}

The authors would like to thank Jonathan Biteau for helpful discussions explaining the PAO methodology. This project has received funding from the European Research Council (ERC) under the European Union's Horizon 2020 research and innovation program (Grant No. 646623).

\section*{Data Availability}
The UHECR propagation simulations performed for this work were done with the software package CRPropa~3~\citep{Batista:2016yrx}, which is publicly available from \url{crpropa.desy.de}. Additional analysis scripts are available on \url{github.com/avvliet/EGMFConstraintsFromUHECRs}.

\bibliographystyle{mnras}
\bibliography{bibliography}

\appendix

\section{Pure-proton scenario}\label{app:protons}

\begin{figure}
	\centering
	\includegraphics[width=0.48\textwidth]{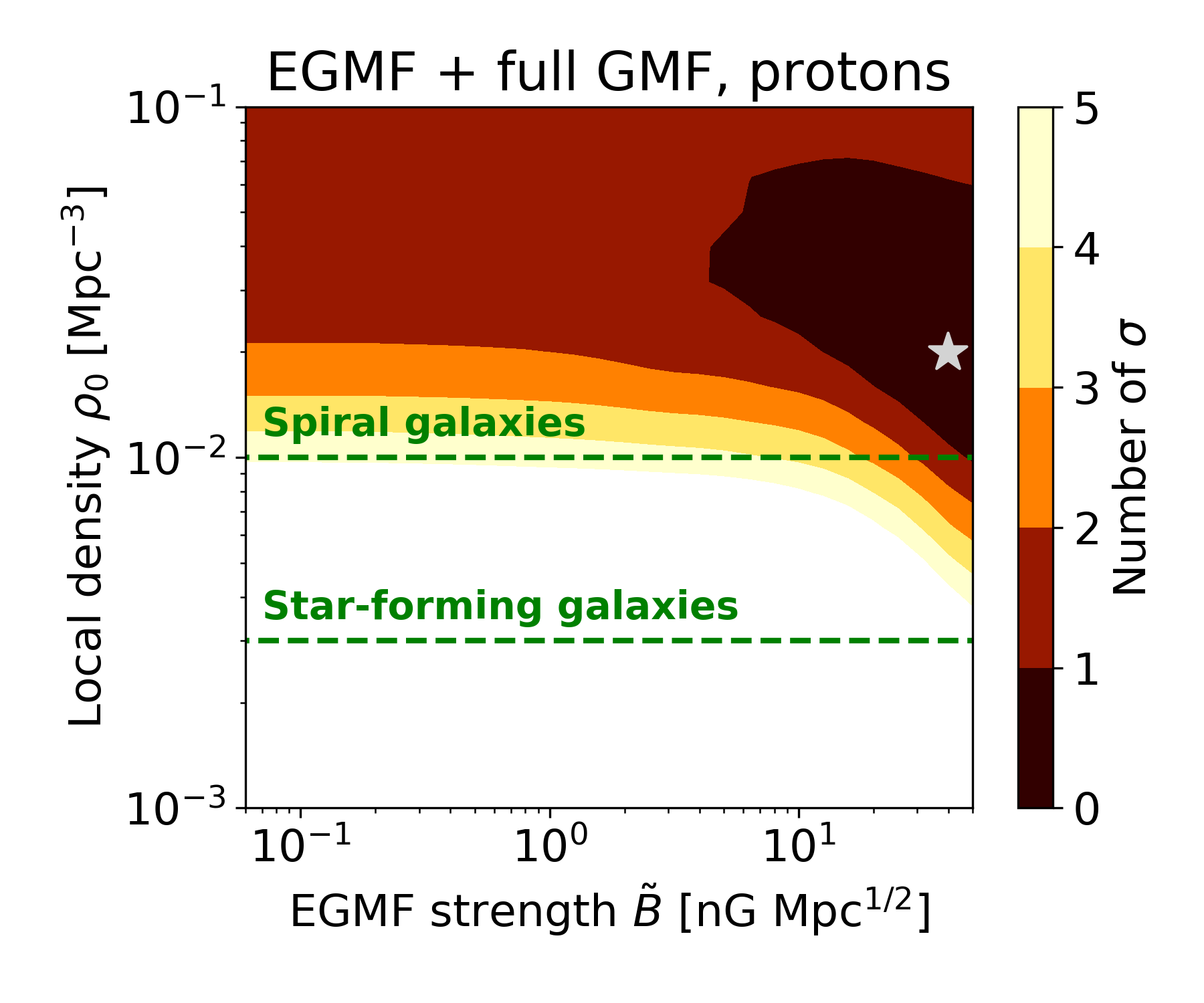}
	\caption{Scenario with extragalactic magnetic field and Galactic magnetic field for a pure-proton composition. We scan over the effective EGMF parameter and over the local density, while the GMF is fixed.  By comparing our results with the results obtained by the PAO, we represent the regions that are $N\sigma$ away from the PAO results. The number of $\sigma$ is obtained using the square root of $\chi^2$ for 2 degrees of freedom. The best-fit point is indicated by the star. We also denote the local densities corresponding to spiral galaxies and to \sbgs~\citep{Gruppioni_2013}. }
	\label{fig:protons}
\end{figure}

We also analyzed the extreme case of a pure-proton composition including both the EGMF and the GMF, Fig.~\ref{fig:protons}. Although this scenario is already excluded by present UHECR composition measurements, it is interesting to analyze an extreme scenario in which the deflection is minimized. For this scenario the spectral index and maximum rigidity at the sources is changed to $\gamma_\text{p} = 2.42$ and $R_\text{max,p} = 10^{21.9}$~V to still obtain a decent fit to the measured UHECR spectrum~\citep{Heinze:2015hhp}.
Assuming a pure-proton composition, the local density should be very large, significantly larger than that of \sbgs, in order to be in agreement with the PAO measurements. The best-fit in this case is located at $\tilde{B} = 40  \ \rm nG \ Mpc^{1/2}$ and $\rho_0 = 2.0 \cdot 10^{-2} \ \rm Mpc^{-3}$. 90\% C.L.~limits on the local source density of $9.0 \cdot 10^{-3} < \rho_0 < 8.4 \cdot 10^{-2} \ \rm Mpc^{-3}$ are obtained.

\bsp	
\label{lastpage}
\end{document}